 \documentclass[manuscript]{aastex}
 \usepackage{wasysym}

\slugcomment{Accepted on 22 November 2011 for publication in ApJ.}

\shorttitle{\sc SO$_{2}$ in VY CMa}
\shortauthors{\sc Fu et al.}

\begin{document}


\title{Arcsecond resolution mapping of Sulfur Dioxide  emission\\ in the circumstellar envelope of VY~Canis~Majoris}


\author{Roger~Fu\altaffilmark{1,2}, Arielle~Moullet\altaffilmark{1}, Nimesh~A.\ Patel\altaffilmark{1,3},\\ John~Biersteker\altaffilmark{1}, Kimberly~L.\ Derose\altaffilmark{1}, Kenneth~H.\ Young\altaffilmark{1}} 

\altaffiltext{1}{Harvard-Smithsonian Center for Astrophysics,  60 Garden
Street, Cambridge, MA}
\altaffiltext{2}{Department of Earth, Atmospheric, and Planetary Sciences,
Massachusetts Institute of Technology, Cambridge, MA}
\altaffiltext{3}{Author for correspondence: npatel@cfa.harvard.edu}

\begin{abstract}
We report Submillimeter Array observations of  SO$_{2}$ emission
in the circumstellar envelope of the red supergiant VY CMa,  with
an angular resolution of  $\approx 1"$. SO$_{2}$ emission appears
in three distinct outflow regions surrounding the central continuum
peak emission that is spatially unresolved. No bipolar structure
is noted in the sources. A fourth source of SO$_{2}$ is identified
as a spherical wind centered at the systemic velocity. We estimate
the SO$_{2}$ column density and rotational temperature assuming
local thermal equilibrium (LTE) as well as perform non-LTE radiative
transfer analysis using RADEX. Column densities of SO$_{2}$ are
found to be $\sim10^{16}$ cm$^{-2}$ in the outflows and in the
spherical wind. Comparison with existing maps of the two parent
species OH and SO shows the SO$_{2}$ distribution to be consistent
with that of OH. The abundance ratio $f_{SO_{2}}/f_{SO}$ is greater
than unity for all radii greater than at least $3\times10^{16}$ cm.
SO$_{2}$ is distributed in fragmented clumps compared to SO, PN,
and SiS molecules. These observations  lend support to specific
models of circumstellar chemistry that predict $f_{SO_{2}}/f_{SO}>1$
and may suggest the role of localized effects such as shocks in the
production of SO$_{2}$ in the circumstellar envelope.
 \end{abstract}


\section{Introduction}

Massive stars ($M_{*}\gtrsim8\,M_{\odot}$), enter a red supergiant (RSG) phase during which the 
star experiences mass-loss at rates of $\dot{M}\sim\,10^{-5}-10^{-3}M_{\odot}yr^{-1}$ \citep{key-18}.
The time variation of this mass-loss rate is not well-constrained by theoretical studies \citep{YoonAndCantiello2010}. As a result, the total amount of mass lost over the course of the RSG phase remains uncertain for a given initial mass \citep{key-20}. Observations of mass-loss events have shown them to be 
sporadic and spatially anisotropic \citep{key-19}.\\
VY Canis Majoris (VY CMa) is an oxygen-rich red supergiant with an estimated mass of $M_{*}\approx 25\,M_{\odot}$ and  a mass-loss rate estimated to be $\dot{M}\sim\,2-4\times10^{-4}M_{\odot}yr^{-1}$ \citep{key-61,key-20}. 
Optical images of this source show multiple discrete and asymmetric mass-loss events, ranging in age
from 1700 to 157 years ago, that are distinct from the general flow of diffuse material \citep{key-31}. 
Detailed studies of mm/sub-mm molecular spectra of VY CMa have been carried out, revealing
the chemical complexity in the envelope  \citep{key-10, RoyerEtAl2010,tenenbaum2010}. Spatial structures in the visible and IR bands  have also been obtained \citep{key-35}. High angular resolution Submillimeter Array (SMA)\footnote{The Submillimeter Array is a joint project between the
Smithsonian Astrophysical Observatory and the Academia Sinica
Institute of Astronomy and Astrophysics, and is funded by the
Smithsonian Institution and the Academia Sinica.} observations 
of VY CMa produced maps of the spatial distribution of CO and SO \citep{key-11}.

Millimeter wavelength observations have long shown evidence of SO$_{2}$ in the circumstellar envelopes (CSE) of oxygen-rich red supergiants  \citep[e.g.,][] {key-62} and SO$_{2}$ emission in VY CMa was first detected by \cite{key-51}. Later infrared observations have also shown the production of SO$_{2}$ within several radii of O-rich AGB stars (in the "inner wind") \cite{key-81}. From a theoretical perspective, the presence of SO$_{2}$ in circumstellar envelopes has been predicted by isotropic, non-equilibrium models of stellar chemistry \citep{key-36, key-1, key-74}. This class of models assume an isotropic geometry of the CSE with specified values for the mass loss rate and expansion velocity. The resulting model CSEs consist of an inner region where assumed parent species in the outflow are broken down, a intermediate region where chemical reactions lead to high abundances of daughter species, and an outer region where the daughter species are destroyed by photodissociation. SO$_{2}$ in these models has been assigned as a daughter species, although it has since been shown to exist as a possible parent species created in the inner wind \citep{key-75}. \cite{key-80} studied the role of shocks in SO$_{2}$ formation in the inner wind within a few stellar radii of the photosphere.

Observational works \citep[e.g.,][]{ jackson1988, key-81,key-75} have repeatedly  shown SO$_{2}$ abundance to be higher both in the extended CSE and the inner wind than expected from CSE and inner wind chemistry models \citep{key-74, key-80}. Local abundance of SO$_{2}$ may be enhanced in the ISM by the passage of shocks  \citep{hartquist1980}, and shock chemistry has been proposed to explain the high observed SO$_{2}$ relative abundances in CSEs \citep{jackson1988}. Alternatively, low model abundances of SO$_{2}$ presented in \cite{key-1} compared to that of \cite{key-36} may be due to the latter's lower assumed value for the photodissociation of SO$_{2}$, which is the main mode of destruction of this molecule at large radii \citep{key-51}. 

Interferometric mapping of SO$_{2}$ distribution in the CSE of an evolved star may contribute to the improved understanding of sulfur chemistry by directly constraining the relative abundances of sulfur-bearing species as a function of radius. Furthermore, associations of SO$_{2}$ enhancement with discrete outflow features may favor the idea of localized production and provide evidence for non-isotropic processes in CSEs \citep{jackson1988}. In this paper we present results of high spatial resolution maps obtained from SMA observations of  SO$_{2}$ around VY CMa. We describe four spatially discrete sources of SO$_{2}$ emission and use multiple transitions of SO$_{2}$ and derive rotational temperatures and column densities (assuming local thermodynamical equilibrium). Furthermore, we perform non-LTE radiative transfer analysis using the RADEX  package \citep{key-70} to derive local kinetic temperatures, H$_{2}$ densities, and SO$_{2}$ abundances of the SO$_{2}$ emitting regions.
We then compare our maps to published visible, IR, and submillimeter observations to evaluate their consistency with the above cited models of circumstellar chemistry.

\section{Observations}

VY CMa was observed with the SMA on 2009 February 18,  with the 
array in extended configuration, offering baselines from 44.2 to 225.9~m.  The projected baseline lengths were from 14 to 188 m.
The frequency coverage was 234.36 to 236.34~GHz in the lower sideband and 244.36
to 246.34~GHz in the upper sideband.
The phase center was at $\alpha(2000)=07^{h}22^{m}58^{s}.27, 
\delta(2000)=-25^{\circ}46'03.4''$. The quasars 0730-116 and 0538-440 were
observed every 20 minutes for gain calibration, and the spectral band-pass was
calibrated using quasar 3c273. Flux calibration was done using recent
SMA measurements of 0730-116 (2.75~Jy at 1~mm) and 0538-440
(3.96~Jy at 1~mm). Nominal flux calibration accuracy is 15 to 20\%, depending on the phase stability.  In our observations, the uncertainty appears to be better than 10\%, based on the good agreement with the ARO spectra (see section 3.2 and Figure 3). The on-source integration time on VY CMa was 5.66~hours.
T$_{sys}$ (SSB) varied approximately from 200 to 400~K during the track with an atmospheric zenith optical depth of $\sim$0.1 at the standard reporting frequency of 225~GHz, measured at the nearby Caltech Submillimeter  Observatory. The conversion factor between Kelvin and Jansky for our observations is $\sim14$ K/Jy. 

The visibility data were calibrated using the MIR package in IDL and imaged with
the MIRIAD software\footnote{http://www.cfa.harvard.edu/$\sim$cqi/mircook.html} \citep{SaultEtAl1995}. The field of view (FWHM primary beam) varies from 53$''$ to 56$''$ whereas the largest angular extent of the source is expected to be about 5$''$. The  synthesized beam size, representing the obtained spatial resolution, is $1.''48\times 1.03''$. 
The adopted weighting mode for imaging is ``natural'' weighting  (specified in the {\it Miriad} task INVERT).
The final rms noise level is $\sim$60~mJy/beam per channel in the spectral line images and 3.9~mJy/beam in the continuum map. Line emission was subtracted b yvisually examining the spectra in visibility amplitudes, and using the line-free channels specification to the {\it Miriad} task UVLIN.

\section{Results}

\subsection{Continuum emission}
We detected  continuum emission in both the upper and lower sidebands, which is well described by a point source centered at RA=$07^{h}22^{m}58.336^{s}$, Dec=$-25^{\circ}46'03.063''$. A 2D Gaussian fit in the image plane yields an integrated flux of $335.0\pm66$~mJy at $235.4$~GHz and $359.3\pm72$~mJy at $245.4$~GHz (maximum errors are of $20\%$, mainly resulting from the uncertainty in flux calibration). Previous SMA observations of VY CMa at $215$ and $225$~GHz have yielded continuum fluxes of $270\pm40$~mJy and $288\pm25$~mJy respectively \citep{key-32,key-11}. These four values are consistent under the Rayleigh-Jeans approximation with black-body radiation with a brightness temperature higher than 12~K. 

\subsection{Line emission}

\begin{figure}
\begin{center}
\includegraphics[width=12cm,angle=-90]{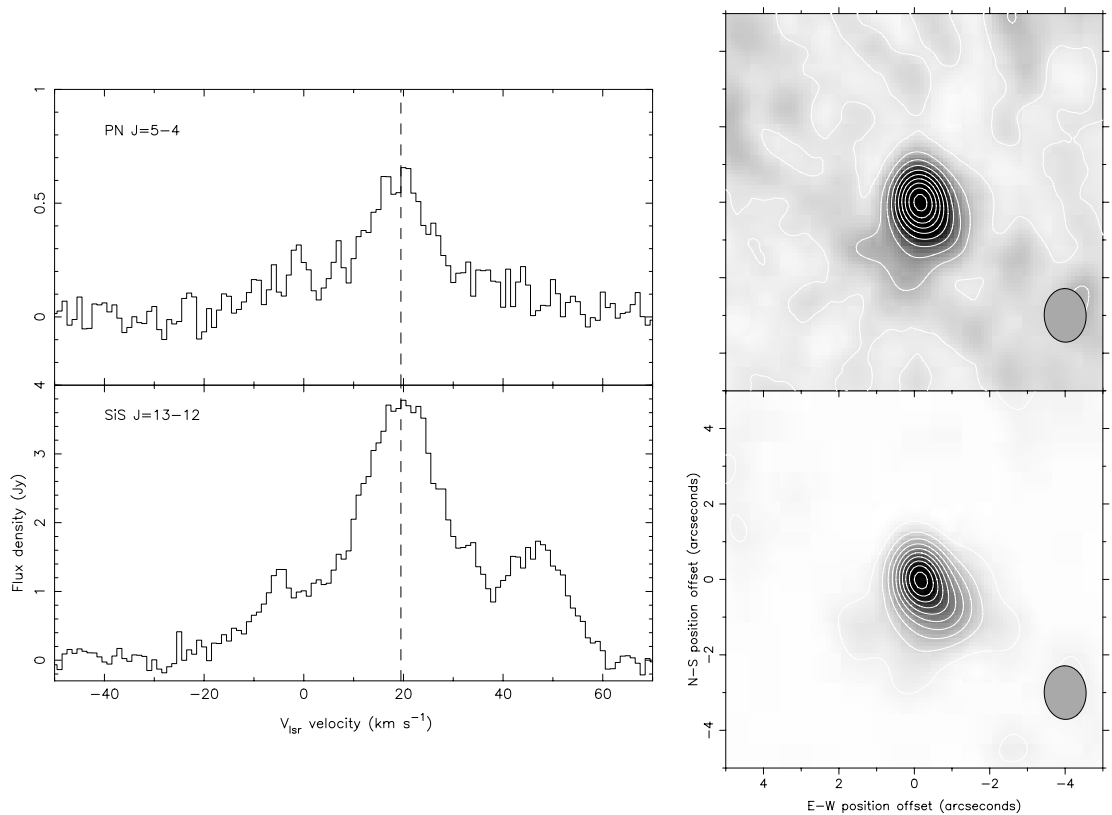}
\end{center}
\caption{ {\it Left:} PN J=5--4 and SiS J=13--12 spectra showing the peak emission at 19.49 km s$^{-1}$ (from the line rest frequencies). We adopt this velocity as that of the stellar frame. ${\it Right:}$ Integrated intensity maps of PN (top) and SiS (bottom) emission,  which show maxima close to the peak of the continuum emission. \label{fig1}}

\end{figure}

In the two sidebands of $\sim$4 GHz bandwidth, we detected a total of 14 lines, which show emissions from CS, H$_{2}$O, PN, SiS and SO$_{2}$. In Figure \ref{fig1} we present the spectra and maps of the SiS J=13-12 and the PN J=5-4 lines. These lines are representative of a relatively compact emission that is centered on the peak of the continuum emission, hence they allow us to estimate the systemic velocity of the star. The SiS spectrum  shows the distinctive triple peak morphology as described by \citet{key-10, tenenbaum2010} that is interpreted as the signature of a slowly expanding shell and a pair of faster outflows nearly collimated with the line of sight. We use the fitted velocity of the central peak of these lines to find the systemic velocity of the star  $v_{LSR}=19.5$ km s$^{-1}$. This value is consistent with other results from mm and sub-mm observations \citep{key-10,key-11} and is adopted as the star rest-frame velocity in the following sections.

\begin{figure}
\begin{center}
\includegraphics[width=12cm]{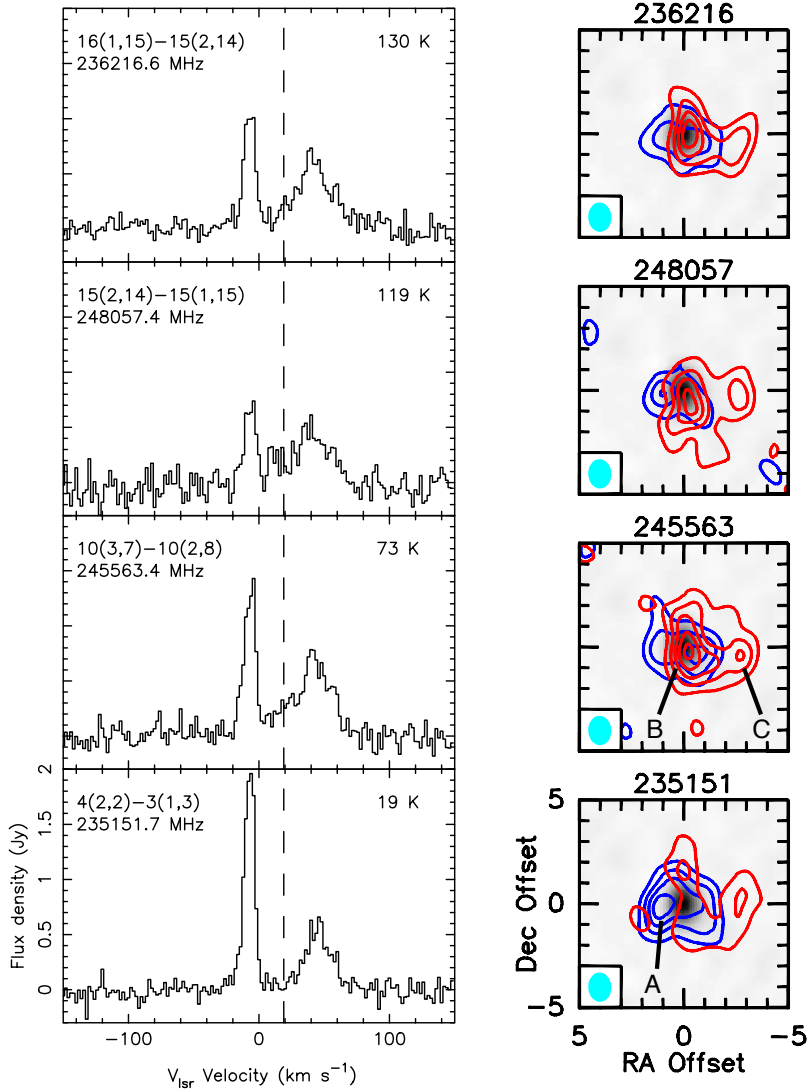}
\end{center}
\caption{ {\it Left:} SO$_{2}$ spectra averaged in a region of 3"x3"
around the continuum peak.  {\it Right:} Integrated intensity maps
corresponding SO$_{2}$ lines over velocity intervals $-20$ and $10$ km s$^{-1}$ are
shown in blue contours. The red contours show the
integrated intensity over velocity intervals 20 to 60 km s$^{-1}$. 
The starting value and interval of contour levels are set to
35 mJy beam$^{-1}$ km s$^{-1}$. The continuum emission is shown in grey scale (repeated
in all panels).  The location of the three identified sources A, B
and C is shown in the bottom two maps.\label{fig2}}

\end{figure}

\begin{figure}
\begin{center}
\includegraphics[width=12cm,angle=-90]{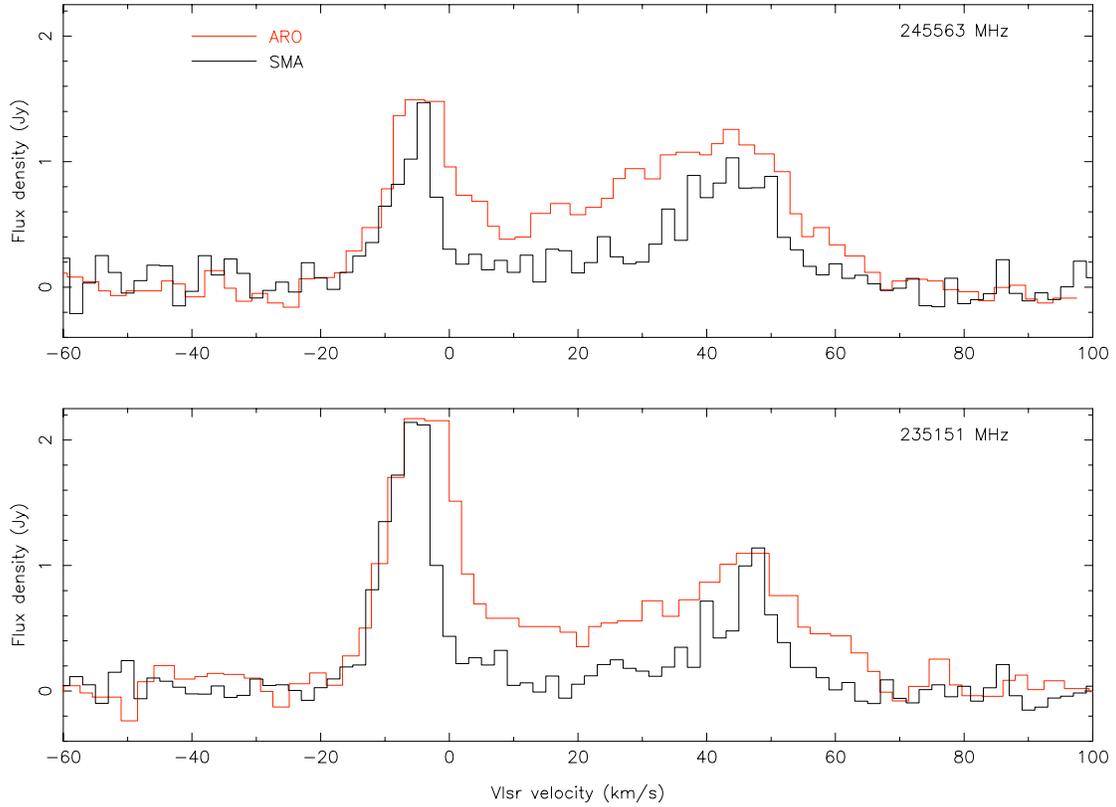}
\end{center}
\caption{ {\it Left:} SO$_{2}$ spectra from our (black) and 10 meter single dish observations by the Arizona Radio Observatory's Submillimeter Telescope (red) for the two lowest temperature transitions (235151.7 and 245563.4 MHz). The SMA data  was smoothed at the appropriate scale to approximate the field of view of the single dish telescope (~30"). \label{fig2.5}}

\end{figure}

\begin{figure}
\begin{center}
\includegraphics[width=12cm]{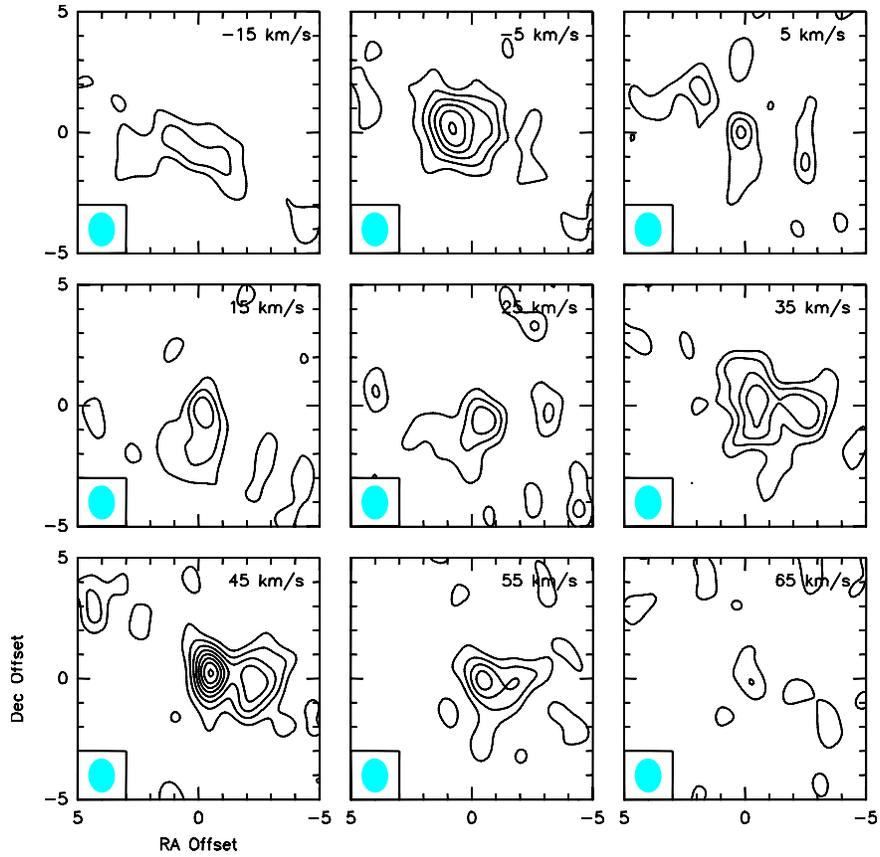}
\end{center}
\caption{Channel maps of SO$_{2}$ 4(2,2)--3(1,3) emission at 235151.7 MHz toward VY CMa.  In each panel, emission is integrated over a 10 km s$^{-1}$ wide velocity range centered on the velocity indicated on the top right corner. Contour levels are same as in Fig. \ref{fig2}. \label{fig3}} 
\end{figure}

\begin{figure}
\begin{center}
\includegraphics[width=12cm]{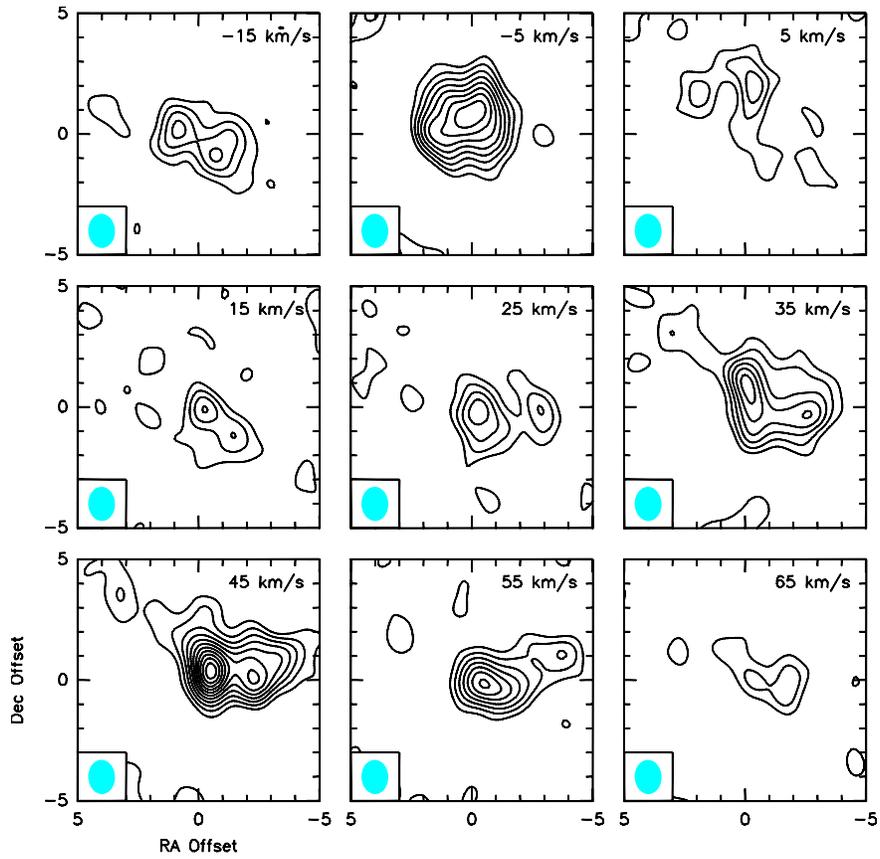}
\end{center}
\caption{Same as in Fig.\ref{fig3} for SO$_{2}$ 10(3,7)--10(2,8) emission at 245563.4 MHz.  \label{fig4}} 
\end{figure}

\begin{figure}
\begin{center}
\includegraphics[width=12cm]{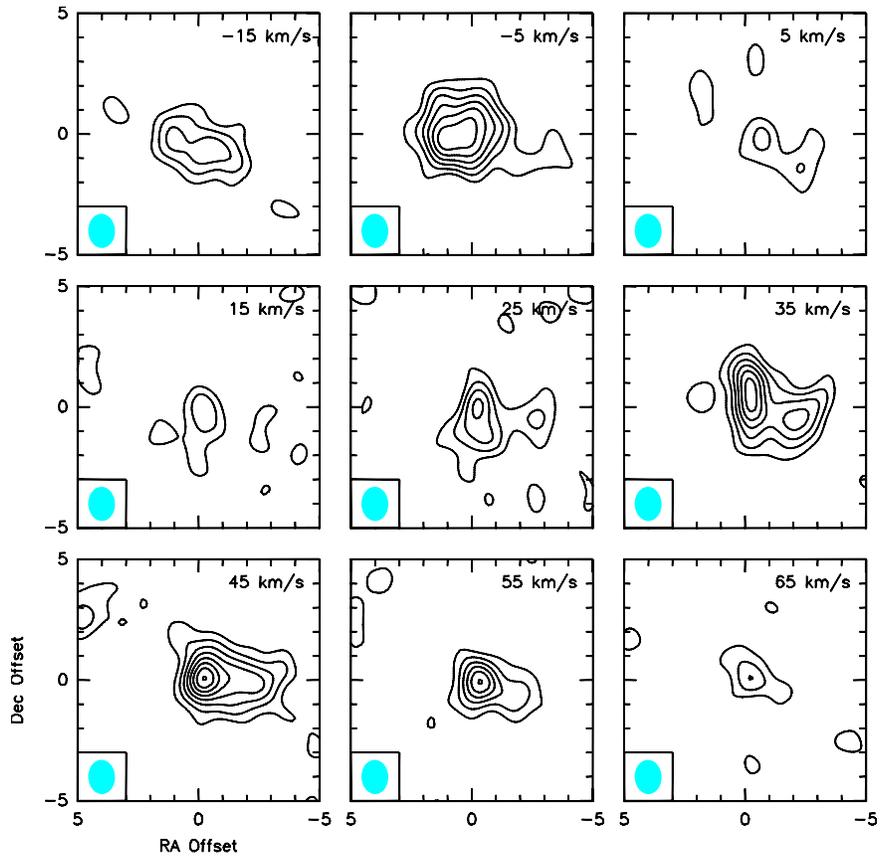}
\end{center}
\caption{Same as in Fig.\ref{fig3} for SO$_{2}$ 15(2,14)--15(1,15) emission at 248057.4 MHz. \label{fig5}} 
\end{figure}

\begin{figure}
\begin{center}
\includegraphics[width=12cm]{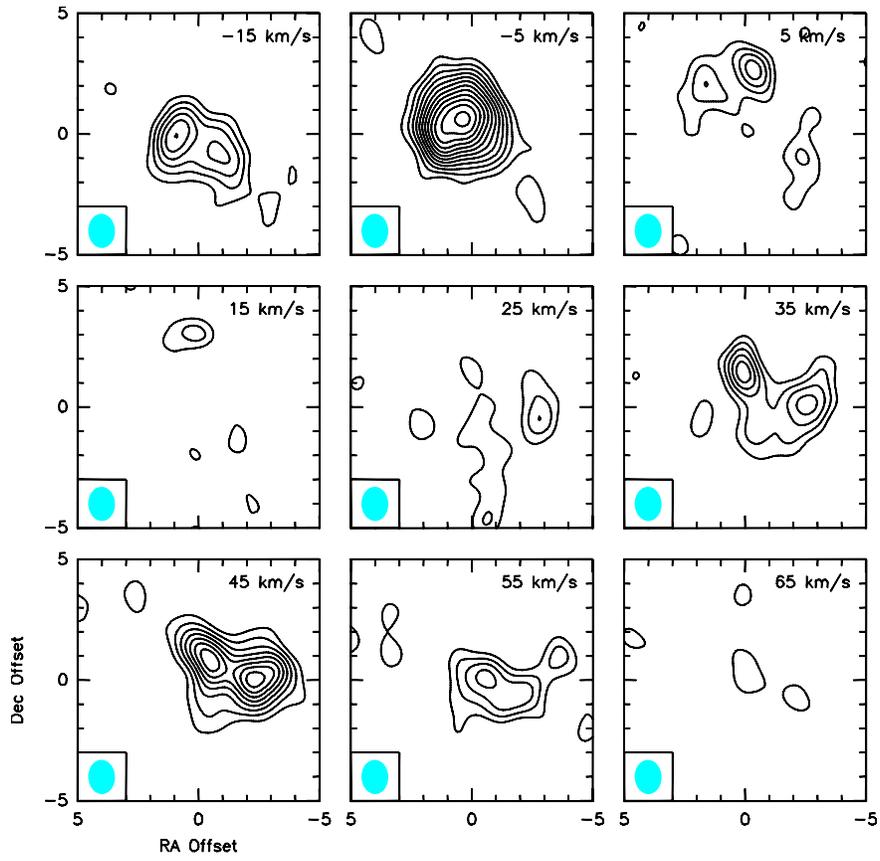}
\end{center}
\caption{Same as in Fig.\ref{fig3} for SO$_{2}$ 15(1,15)--15(2,14) emission at 236216.6 MHz.  \label{fig6}} 
\end{figure}

For the remainder of this work we focus only on the SO$_{2}$ lines,  which are summarized in Table \ref{table1}.  The excitation energies of the four identified SO$_2$ lines vary from 19 to 130 K. All four of these lines were detected in the spectral line survey of \cite{tenenbaum2010}. Our interferometric observations reveal the spatial distribution of the SO$_{2}$ emission for the first time.  The spectra and integrated maps of all four lines are shown in Figure \ref{fig2}. Figure \ref{fig2.5} compares the spectra of two SO$_{2}$ lines with single dish data from the Submillimeter Telescope of the Arizona Radio Observatory, while channel maps of all four lines are presented in Figures \ref{fig3}, \ref{fig4}, \ref{fig5} and \ref{fig6}.

\begin{figure}
\begin{center}
\includegraphics[width=12cm,angle=-90]{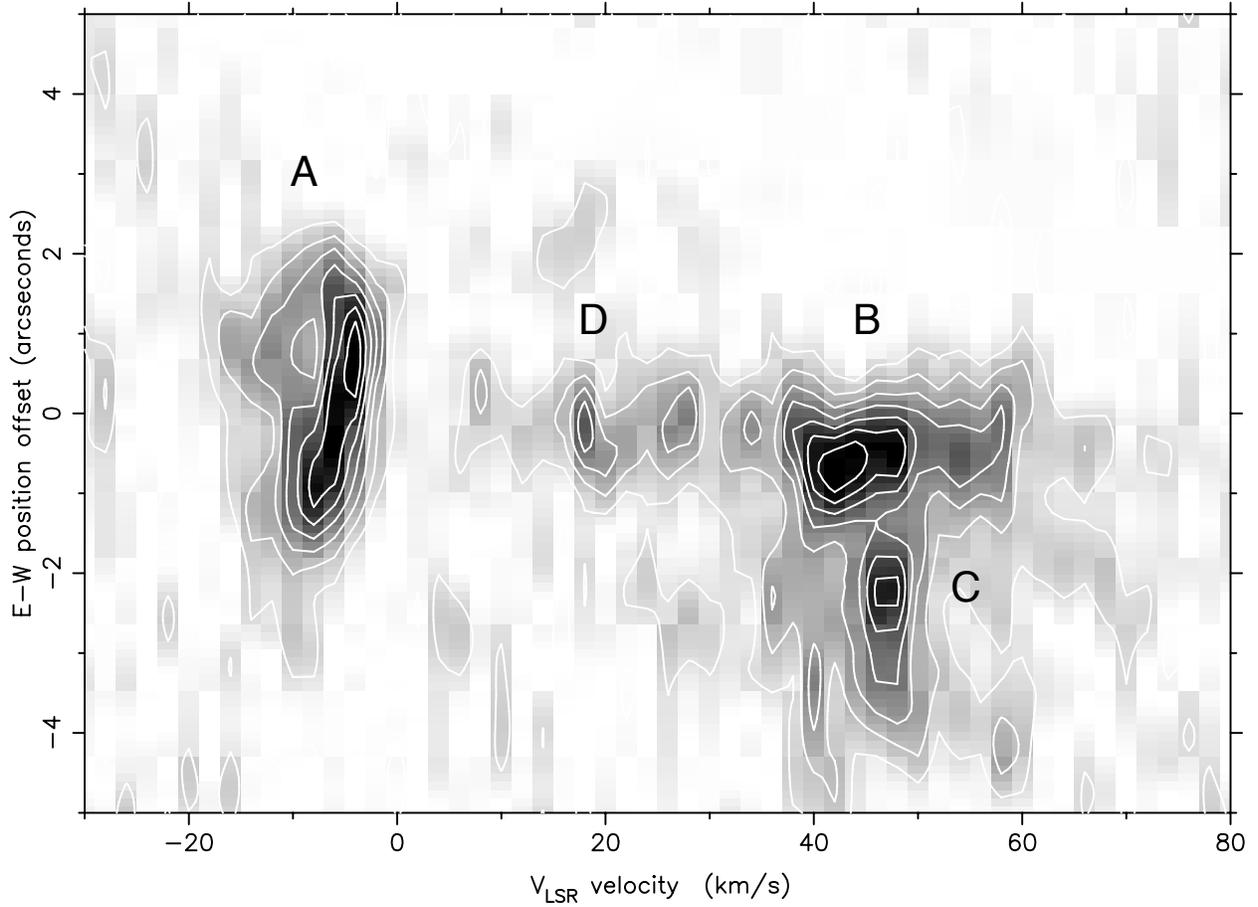}
\end{center}
\caption{ {\it Left:} A position velocity cut along east-west at Declination offset of 0$''$ in the SO$_{2}$ line emission at 245563.4 MHz, at position angle 105$^{\circ}$. The contours correspond to 10\% of the peak emission. Sources A, B, C and D are clearly separated.\label{fig7}}
\end{figure}

Maps of SO$_{2}$ lines maps  show well resolved  spatially extended emission over an area of $3'' \times 5''$, in which we distinguish four distinct components, hereafter designated as sources A, B, C, and D. Figure \ref{fig7} shows all four sources in a position-velocity space diagram of the 245563.4 MHz line. Source positions, orientations, and intensities were fitted to 2D Gaussians using the MIRIAD routine IMFIT.  Source A is a very strong blue-shifted source with $v_{lsr}$  between $-18$ and $+10$~km s$^{-1}$. It is spatially offset to the East of the stellar position by $\sim0.7"$. Of similar intensity to source A is source B, an elongated source offset by $\sim0.8"$ to the West. Source B has a broader spectral profile than source A and is heavily red-shifted with $v_{lsr}$ between $+26$ and $+66$ km s$^{-1}$. Source C is a strong source with a similar radial velocity as source B. In contrast to sources A and B, however, it is heavily offset to the West by $\sim2.5"$. 

Source D is significantly weaker than sources A, B, and C. In velocity space, it is centered on the star's frame of motion ($12$ km s$^{-1}$ $<v_{LSR}$$<26$ km s$^{-1}$) with a distinctly narrower spectral profile and similar spatial dimensions as the other sources. It is also spatially centered at the star's location. 

Assuming local thermal equilibrium, we can use the integrated intensities measured in the detected SO$_{2}$ lines in each of the four identified SO$_{2}$ sources to constrain their respective column density $N$ and rotational temperature $T_{rot}$, using the expression from   \cite{key-33}:

\begin{equation}
\ln[\frac{3c^{2}I_{beam}}{16\pi^{3}B\theta_{a}\theta_{b}\nu^{3}S\mu^{2}}]=\ln[\frac{N}{Z}]-\frac{E_{u}}{T_{rot}}\label{eq:lnln}\end{equation}

where $I_{beam}$ is the integrated flux in Jy km s$^{-1}$ beam$^{-1}$,
$\theta_{a}\theta_{b}$ is the beam size, and $B$ the beam filling factor
varying from $0.8$ to $0.95$ for the four distinct sources, as derived from \cite{key-50}:
$\Theta_{source}/(\Theta_{beam}+\Theta_{source})\label{eq:fillfactor}$. $Z$ is the rotational partition function \citep{key-1000}, $S$ is the line strength, $\mu$ the dipole moment (1.63 Debye for SO$_{2}$),
$E_{u}$ the line's upper rotational state energy in K taken from the JPL spectroscopic database \citep{key-53}, and $\Theta$ the solid angle. 

\begin{figure}
\begin{center}
\includegraphics[width=12cm,angle=-90]{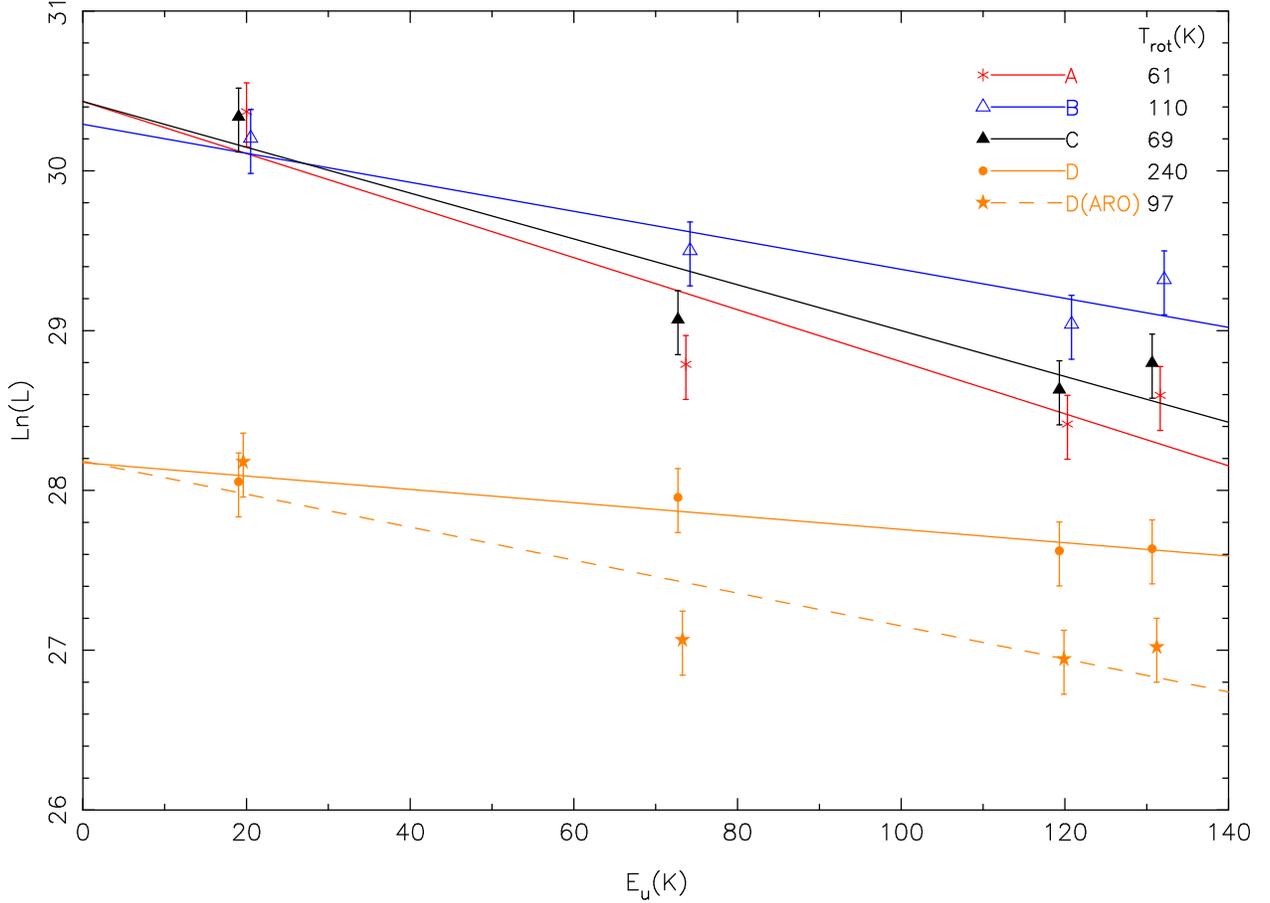}

\caption{Rotational temperature diagram of the four SO$_2$ emission sources. Red, blue, and black curves and points indicate sources
A, B, and C respectively. Orange represents the spherical wind (source D). Variable ``L'' on the vertical axis represents the left-hand side of Equation \ref{eq:lnln}. The dashed orange line represents the spherical wind based on ARO single dish data \cite{tenenbaum2010}. The errors plotted are set to three sigma. Horizontal coordinates of the plotted points have been staggered for clarity\label{fig8}.}
\end{center}

\end{figure}

By performing linear fits of the rotational temperature diagram shown in Figure \ref{fig8}, we find that the three faster outflows (sources A, B, and C) show variations in rotational temperature. Sources A, B, and C have rotational temperature of $61_{-17}^{+39}$ K (one $\sigma$ error), $110_{-29}^{+63}$ K, and $69_{-17}^{+34}$ K, respectively. The column densities (N) of all three sources are similar (A: $9.2_{-4.4}^{+7.8}\times10^{15}$~cm$^{-2}$; B: $1.9_{-0.6}^{+0.8}\times10^{16}$~cm$^{-2}$; C: $1.1_{-0.4}^{+0.7}\times10^{16}$~cm$^{-2}$). The rotational temperature of the close-in source D is much higher at $240_{-52}^{+91}$ K, while its column density is likely lower than that of the others at $7.6\pm1.0\times10^{15}$~cm$^{-2}$. 

The derived parameters for source D are uncertain since a significant portion of flux in its velocity range was missed by our interferometric observations due to the lack of short spacing data (our shortest projected baseline 14 m). Comparison with single dish data from the Arizona Radio Observatory shows that features as large as 4" are fully recovered (Figure \ref{fig2.5}). Features much larger than 4" may be subject to an underestimation of its flux, although 4" is an highly conservative estimate. To address this deficiency, we repeated the rotational temperature diagram analysis for source D using a single dish observation made by the Arizona Radio Observatory 10 meter Submillimeter Telescope \citep{tenenbaum2010}. Assuming a source size of 4" by 4" we derive a lower rotational temperature of $97_{-31}^{+84}$ and a column density of $1.9^{+2.9}_{-1.8}\times10^{15}$ cm$^{-2}$. Because this column density represents an average over our assumed source size, it cannot be directly compared to our derived column densities, which represent the values at the center of each source. This averaged column density also represents an upper limit due to our assumption of the smallest possible angular size. 

The validity of the LTE approximation can be assessed by checking the linearity of the data points in the rotational temperature diagram. Deviation from a linear trend suggests non-LTE conditions in the source or line misidentification. In the case of source D, deviation from linearity is very small compared to the error. We are therefore confident that LTE is a valid assumption for this source. For the outflow sources A through C, a systematic concave up shape is noted in the rotation temperature diagram, suggesting departure from LTE conditions. 

These findings justify the need to consider a non-LTE radiative transfer model to interpret the measured fluxes. We perform this analysis using the RADEX package \citep{key-70}, which calculates expected line intensities for a given column density of SO$_{2}$, kinetic temperature (T$_{kin}$), and volume density of H$_{2}$. For each source, we assume spatially homogeneous kinetic temperature and volume density. The SO$_{2}$ column density is fixed for each source and its value is drawn from the LTE analysis, which gives $10^{16}$ cm$^{-2}$ for sources A, B, and C and $10^{15}$ cm$^{-2}$ for source D. Varying column density one order of magnitude in each  direction does not significantly affect the results. We perform calculations for a wide range of values for the kinetic temperature and the H$_{2}$ density. We then evaluated the resulting ratios between the flux of each line and that of the $235151.7$ MHz transition for the outflow sources A, B, and C. Due to the lack of emission from the $235151.7$ MHz line in the vicinity of source D, the 236216.7 MHz line was used instead to calculate line ratios. 

\begin{figure}
\begin{center}
\includegraphics[width=12cm,angle=90]{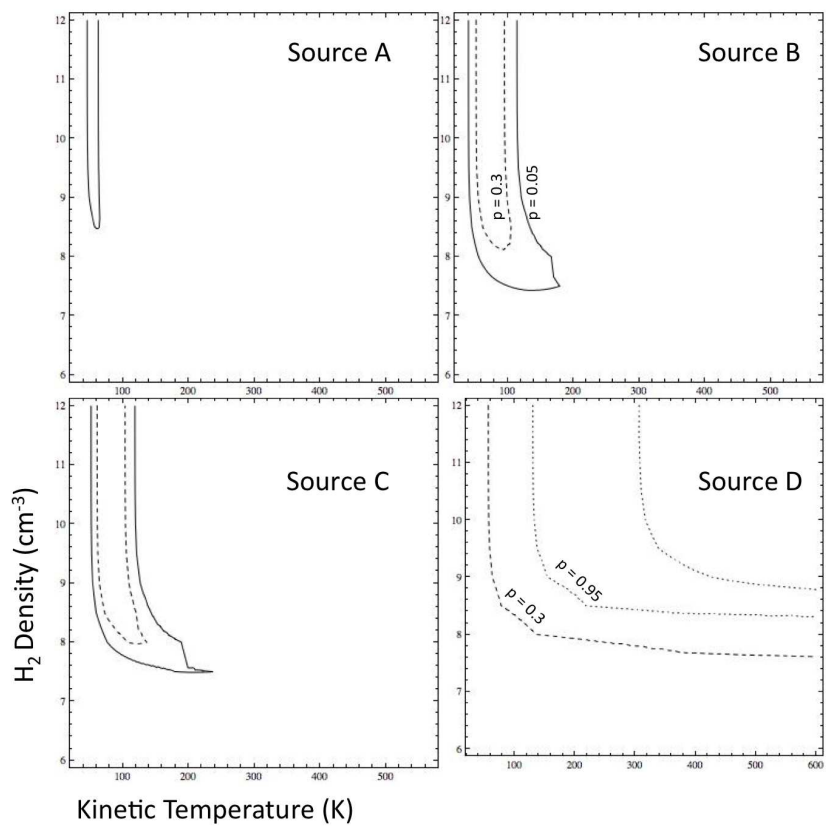}

\caption{Fits between our measured line ratios and predictions with Radex radiative transfer. Solid, dashed, and dotted contours enclose regions of $\chi^{2}$ test p-values greater than 5\%, 30\%, and 95\%. Owing to lower S/N and weak dependence of line ratios to these quantities, the derived temperature and density of source D are much less well-constrained than the other three. \label{fig9}}
\end{center}

\end{figure}

Fits to our data are presented as the $\chi^{2}$ statistic p-value for each assumed value of T$_{kin}$, and H$_{2}$ density (Figure \ref{fig9}) and best fit results are tabulated in Table \ref{table3}. We see that the non-LTE RADEX results for $T_{kin}$ show general agreement with $T_{rot}$ derived with LTE assumption. A reliable estimate of H$_{2}$ density is elusive. Because the flux of each line becomes more similar at greater values of H$_{2}$ density, we are able only to constrain a lower bound on this value. 

The constraints on the temperature and H$_{2}$ density in source D are much weaker. However, the best fit values of T$_{kin}$ span the 130 K to 330 K range, which brackets the value of T$_{rot}=240$ K obtained from the LTE analysis. This agreement between T$_{rot}$ and T$_{kin}$ is consistent with LTE conditions in source D. Our RADEX analysis also allows us to evaluate our assumption of optically thin SO$_{2}$ lines necessary for the LTE analysis performed above.  Although for lower H$_{2}$ densities ($\apprle 10^7$ cm$^{-3}$), the 235.151 GHz line comes close to being optically thick, for the inferred H$_{2}$ densities, all four SO$_{2}$ have optical thickness below $10^{-2}$.

Assuming that the extent of the sources in the line of sight direction is similar to that in the plane of the sky, and that the density is homogenous, we estimate the fractional abundance of SO$_{2}$,  $f_{SO_{2}}$ in each source. These are tabulated  in Table \ref{table3}.

\section{Discussion}
\subsection{Source properties}
We begin by discussing the positions of the four identified SO$_{2}$ sources. Source C, offset $2.5"$ to the West, is clearly an isolated body with no antipodal companion to the East. Its existence was suspected in \cite{key-10}, but was not treated as a distinct source. On the other hand sources A and B are found to be offset in opposite directions relative to the star position with similar blue and redshift velocities relative to the stellar frame. SMA observations of the CO and SO analogs of sources A and B were interpreted to be antipodal companions oriented $15^{\circ}$ from the line of sight by \cite{key-11}. On the other hand, \cite{key-12} found that single dish observations of CO and other molecules are best explained by a blueshifted and a redshifted source at 20$^{\circ}$ and 45$^{\circ}$ from the line of sight. Similarly, visible and IR HST observations \citep{key-31, key-35} have found no evidence of antipodal structure around VY CMa. A careful inspection of sources A and B in our observation shows that the faster (in radial velocity relative to the star's reference frame) sections of both bodies are offset towards the southwest, while the slower sections are offset towards the northeast. This observation argues against a bipolar geometry, in which antipodal subsections of the two outflows should have similar velocities. We therefore treat sources A, B, and C as three distinct outflows probably unrelated to the symmetry axis of the star. These sources do not seem to correspond to any visible or IR features. Our maps show that all SO$_{2}$ outflow sources are too close to the star to be identified as the "curved nebulous tail" or the numbered arcs presented in \cite{key-35}, which are located $\sim3.5"$ from the star.

We identify sources A and B with the blue and redshifted outflows of the previous authors \citep{key-10,key-11,key-12}. The P-V space morphology of these sources match closely with outflows of both CO and SO lines mapped by \cite{key-11} (Figure \ref{fig7}). These bodies are hypothesized to have originated in an episode of anomalously high mass loss at uncorrelated locations on the stellar surface \citep{key-35}. Assuming that the ages of the outflows are similar and adopting the $\sim500$ year age found by \cite{key-11}, we can attempt to derive their locations in three-dimensions. We find that the deprojected radii of sources A and B are $4.2\times10^{16}$ and $4.8\times10^{16}$ cm and that both are situated at $22^{\circ}$ from the line of sight. Their deprojected star-frame velocities of $28$ and $30$ km s$^{-1}$ fall within the range of measured outflow velocities from the multi-epoch observations of \cite{key-31}. If we assume the same age as for sources A and B, then source C is found at a similar radius from the star ($6.9\times10^{16}$ cm), but it is much faster at $44$ km s$^{-1}$ and is situated $54^{\circ}$ from the line of sight. SO$_{2}$ abundance at large radii is expected to be controlled by a balance between the rates of production, expansion, and photodissociation \citep{key-36}. Outflows with faster expansion velocity are expected to maintain high SO$_{2}$ abundance out to greater radii, as in the case of our source C.

Source D is elongated and centered at the stellar position,  with a resolved minor and major radii of $1"$ and $1.6"$, corresponding to $2.2\times10^{16}$ and $3.5\times10^{16}$ cm. The minor axis radius corresponds to between $180$ and $540$ stellar radii, depending on the adopted stellar radius \citep{key-71, key-72}. A significant proportion of source D flux is missing from our observation when compared to single dish results (Figure \ref{fig2.5}). The source we observe therefore appears to represent the warm core of a larger extended envelope found at the systemic velocity with lower average column density.  This extended source is analogous to the spherical wind described in previous millimeter wavelength observations of VY CMa by \cite{key-10}, \cite{key-12}, \cite{key-11}, and \cite{tenenbaum2010}; however,  only the first and last of these works observed SO$_{2}$ and did not identify  it in the spherical wind. These previous works have found a relatively low expansion velocity of between $15$ and $20$ km s$^{-1}$, which is high given the narrow velocity range of source in our channel maps (Figures \ref{fig3} - \ref{fig6}). However, this discrepancy may be due to missing source D flux in our observations. The rotational temperatures derived from SMA and ARO data indicate the existence of a thermalized compact region with elevated temperatures with diameter $\approx 3\times10^{16}$ cm. In the sections below, we refer to this inner region as the core of source D.

Our derived values of $T_{rot}$ and $T_{kin}$ distinguish between the lower temperatures of sources A, B, and C and a much hotter core of source D. In comparison to previous works, our temperatures for source A, B, and C bracket the range of temperatures derived by \cite{key-11} and \cite{key-12} ($57$K and $85$K, respectively). This may be expected, as the preceding authors adopted the same best fit power-law temperature profile for both outflows, which are assumed to be at the same radius. As such, these previous results may represent an average of the temperatures of the outflows. 

Our inferred H$_{2}$ densities from RADEX radiative transfer analysis are  higher than the values found in previous studies. \cite{key-12} adopted an isotropic H$_{2}$ density profile based on an assumed mass-loss rate that gives a value of $\sim1\times10^{6}$ cm$^{-3}$ at $10^{16}$ cm radius. \cite{key-11} use a similar procedure to arrive at a lower value of $4.5\times10^{5}$ cm$^{-3}$ at the same location.  A higher density of $3\times10^{6}$ cm$^{-3}$ is assumed for the inner wind region in \cite{tenenbaum}. 

Part of this discrepancy between our values for H$_{2}$ density and that of previous authors may be explained by the latter's assumption of isotropic mass flow, which does not account for density concentration in the spatially confined outflow regions. However, this reason alone may not be able to account for the more than two orders of magnitude difference. More significantly, our derived densities of between $5\times10^{6}$ and $2\times10^{8}$ cm$^{-3}$ may be due to the presence of SO$_{2}$ in regions of local density enhancement above the expected values from an isotropic model. We speculate that such high density regions are the result of shocks, and their existence around VY CMa can be inferred from the observations of OH masers, which overlap with sources A and B of SO$_{2}$ emission \citep{key-37}. The activation of the observed 1612 MHz maser line requires H$_{2}$ densities of between $1\times10^{6}$ and $3\times10^{7}$ cm$^{-3}$ at 100 K \citep{Pavlakis1996}. The upper end of this range is similar to our inferred value of minimum H$_{2}$ density for sources B and C, while that of source A is much higher. However, OH masers may still be active in our source A despite its high density since its temperature of $\sim55$K is cooler than that assumed in \cite{Pavlakis1996}.

\subsection{Sulfur chemistry in the CSE}
For the remainder of the Discussion, we address the  implications of our observations for circumstellar chemistry by comparing SO$_{2}$ distribution with those of the OH and SO molecules. SO$_{2}$ in CSEs is formed via the following reaction \citep{key-36, key-1, key-74, key-80}: 

\[
SO+OH\longrightarrow SO_{2}+H\]

The radial abundance of SO$_{2}$ is therefore expected to reflect that of the two reactant molecules. A striking similarity between the spectral profiles of SO$_{2}$ and OH masers has already been noted by \cite{key-10}. Maps of the 1612 MHz OH maser line show that it coincides with SO$_{2}$ in sources A and B, but it is weak or undetectable in the outflow source C or the spherical wind source D. The lack of OH maser detection in source C (or perhaps a very weak detection; see \citealp{key-37}) may be explained by anisotropic nature of maser radiation. Assuming that the velocity of outflows is oriented radially away from the star, relative velocities between different clumps of gas along a photon's line of travel are smallest when the path is parallel or antiparallel to the gas expansion velocity. The most efficient pumping of a masering state is then achieved when photons travel radially inward or outward from the star. Therefore, the strongest maser emissions are observed from sources along our line of sight \citep{elitzur}. This condition is nearly met for sources A and B ($15^{\circ}$ to $22^{\circ}$ from the line of sight). On the other hand source C, found at a line of sight angle of $54^{\circ}$, may also produce the 1612 MHz OH maser, but its signal is weak along the line of sight.

The weak OH maser emission in the stellar velocity frame may be attributed to the high kinetic temperature of gas in this region. For a given volume density of gas, temperatures above a certain threshold tend to induce thermal equilibrium in the emitting body, undoing the population inversions responsible for maser emissions  \citep{Pavlakis1996}. For temperatures of $\sim200$K, the maximum allowable H$_{2}$ density for the 1612 MHz OH maser is $3\times10^{6}$ cm$^{-3}$, which is more than an order of magnitude lower than our inferred density for source D. The highly linear trend of Source D data in the rotational temperature diagram (Figure \ref{fig8}) corroborates the prevalence of LTE conditions in source D.

We therefore find that OH and SO$_{2}$ distributions are generally similar and that their differences are reconcilable. 

Finally, we address the discrepancies between the SO$_{2}$ and SO distributions of \cite{key-11}, who have mapped the distribution of SO around VY CMa at a similar resolution to ours using a single rotational line of SO (J = 6$_{5}$ - 5$_{4}$; $E_{u}=35$K). SO was found in all four source regions described in this work, although the red-shifted SO emitter is not fragmented into two discrete sources as in the case of SO$_{2}$. Because of the strongly contrasting values of H$_{2}$ density adopted in \cite{key-11} and this work, direct comparisons between column densities are more instructive than comparisons between fractional abundances. For column density to act as a valid proxy for abundance, we must assume that the line-of-sight dimension of the corresponding  SO and SO$_{2}$ sources are similar and that the distribution of the each molecules within each source is homogeneous. While we cannot be certain that the line-of-sight thickness of the SO and SO$_{2}$ sources are equal, their plane of sky dimensions are similar ($\sim$30\% discrepancies). The combined column density of both SO outflows and the inner wind where they overlap along the line of sight was found to be $\sim10^{16}$ cm$^{-2}$. This value reflected the combined column density from both outflows and the spherical wind (corresponding to our sources A, B, and D). Given that the three regions contribute approximately equal amounts to the total SO column density, the value of $N_{SO}$ in each region is on the order of a few $10^{15}$ cm$^{-2}$. In contrast, we find column density of SO$_{2}$ to be $1$ to $2\times10^{16}$ cm$^{-2}$ in $\it{each}$ outflow source. The column density of SO$_{2}$ in the outflow sources A and C is therefore $\apprge3$ times greater than that of SO in the same regions. This statement likely applies as well to outflow source C, given its similar SO$_{2}$ column density compared to the other outflows and the lack of a discrete SO source at its location. Furthermore, even with missing flux, the column density of SO$_{2}$ in the compact core of source D is within error as that of the outflow sources, implying that $N_{SO_{2}} > N_{SO}$ in the core of the spherical wind $ < 3\times10^{16}$ cm from the star.

Non-equilibrium CSE chemistry models \citep{key-36, key-1, key-74} make differing predictions about ratio of SO$_{2}$ to SO abundance ($f_{SO_{2}}/f_{SO}$). 
The \cite{key-36} (SS) model adopts a reaction rate for SO$_{2}$ formation that is fast compared to the rate of SO$_{2}$ destruction via photodissociation. Under these conditions, SO is quickly converted into SO$_{2}$, which becomes the primary S-bearing species. SO$_{2}$ therefore reaches maximum abundance at a greater radius than SO. SS has predicted the radius of this transition where $f_{SO_{2}}/f_{SO}>1$ to be around several times $10^{15}$ cm. 

Our observations suggest that $f_{SO_{2}}/f_{SO}>1$ for the outflow regions ($>4\times10^{16}$ cm from star) and in the compact region within $3\times10^{16}$ cm of the star. If there does exist a transitional radius at which $f_{SO_{2}}/f_{SO}=1$, it must be less than the latter radius, making our study consistent with the \cite{key-36} model. On the other hand, models that predict $f_{SO_{2}}/f_{SO}<1$ at all radii are inconsistent with our data \citep{key-1,key-74}. In addition, the SS model predicts a steady increase in the value of $f_{SO_{2}}/f_{SO}$ outward from the $f_{SO_{2}}/f_{SO}=1$ transition radius at $\sim2\times10^{15}$ cm. The value of $f_{SO_{2}}/f_{SO}$ increases by more than an order of magnitude by radius 10$^{16}$ cm. Observations of the outflow sources A, B, and C do not support this view, as $f_{SO_{2}}/f_{SO}$ in on the order of ~3 for all four sources. More precise comparisons between SO$_{2}$ and SO abundance in each source is hindered by the lack of more detailed interpretations of SO maps (the \cite{key-11} observation included only one SO line).  Therefore, quantitative comparison between model radial distribution of the two species and our results is elusive and we regard our support of the SS model as only qualitative.

Other uncertainties remain in our understanding of sulfur chemistry in the CSE of VY CMa. Our observations cannot be used to constrain whether the SO$_{2}$ originates in the CSE, as assumed by the cited models, or in shocks in an "inner wind" within a few stellar radii of the photosphere as modeled by \cite{key-80} and observed by \cite{key-75} and \cite{key-81}. Furthermore, as proposed earlier by \cite{jackson1988} and \cite{key-74}, shock events similar to those that occur in the inner wind may be the source of SO$_{2}$ enrichment in the outflow lobes. Indeed, comparison of SO$_{2}$ and maps of SO, PN, and SiS (\citealp{key-11}; Figure \ref{fig1}, this work) shows that SO$_{2}$ exhibits a more fragmented distribution with discrete sources in the red-shifted outflow. Unlike these other molecules, the red-shifted SO$_{2}$ outflow is partitioned into two regions of high local abundance (sources B and C), suggesting that local effects may participate in the formation of SO$_{2}$. However, CSE chemistry models that include the effect of shocks (similar to \citealp{key-80}) remain to be done. 

A further uncertainty involves the very high mass loss rate of VY CMa, which is, for example, about two orders of magnitude faster than the asymptotic giant branch star IK Tau \citep{key-1001}. Varying the mass loss rate in CSE chemistry models does not significantly affect the radial abundance profiles of chemicals species while it does result in globally larger envelopes for higher mass loss rates \cite{key-74}. Therefore the discussions of the $f_{SO_{2}}/f_{SO}$ profile in this work are also relevant to stars with lower mass loss rates, while the actual measured radii of peak SO$_{2}$ abundance around VY CMa are unique this object. 

\subsection{Summary}
\begin{enumerate}
\item Four rotational lines of the SO$_{2}$ molecule are mapped with $\sim1"$ resolution around VY CMa. SO$_{2}$ is found in four discrete sources, three of which are fast ($28$ to $44$ km s$^{-1}$) outflows far from the star and one is a slower spherical wind near the star. No symmetrical relationship among the faster outflows or visible and IR features are found.
\item The three fast outflows are found at similar distances from the star and  probably originated around 500 years ago.
\item Comparison between our SO$_{2}$ maps and those of the 1612 MHz OH maser line suggests that the two species are strongly correlated and that the OH maser detection may be limited by high temperature and density in the spherical wind.
\item SO$_{2}$ is more abundant than SO in all three outflow sources, supporting the non-equilibrium chemistry model of \cite{key-36}. It is inconsistent with models that predict $f_{SO_{2}}/f_{SO}<1$ for all radii (e.g., \citealp{key-1,key-74}). The distribution of SO$_{2}$ in discrete clumps when compared to other molecules may point to the role of localized effects, such as shocks, in the enhancement of SO$_{2}$ abundance.
\end{enumerate}

\acknowledgements
We are grateful to Raymond Blundell, Thomas Dame and Patrick Thaddeus for the opportunity to carry out this  research using the SMA as part of the Laboratory Astrophysics (Astro 191)  course at Harvard University. We also thank Carl Gottlieb for helpful comments on the manuscript.

\begin{table}

\caption{Summary of observed SO$_{2}$ lines. \label{table1}} 

\begin{tabular}{lllr}

\hline
\hline
SO$_{2}$  & Frequency & E$_{u}$\tablenotemark{1} & $S$\tablenotemark{2} \\
transition & (MHz) &(K) & \\
\hline
$4_{2,2}-3_{1,3}$         & 235151.7 & 19.0 & 1.71 \\
$10_{3,7}-10_{2,8}$     & 245563.4 & 72.7  & 5.44 \\ 
$15_{2,14}-15_{1,15}$ & 248057.4 & 119.3  &  5.27\\
$16_{1,15}-15_{2,14}$ & 236216.7 & 130.7  & 6.05 \\
\hline
\end{tabular}
\tablenotetext{1}{\bf E$_{u}$ is the upper rotational state energy.}
\tablenotetext{2}{\bf $S$ is the line strength.}
\end{table}

\vspace{2cm}

\begin{table}
%\rotate
\caption{Description of SO$_{2}$ line emission sources\tablenotemark{*}\label{table2}}
%\tablewidth{0pt}
\small
\begin{tabular}{lcccc}
\hline
\hline
Source  & Offset  & Position & Size  & Orientation  \\
 (v$_{lsr}$ range in km s$^{-1}$) &  ({}``)  & Angle ($^{\circ}$) & ({}``) & \\
\hline
A: ($-18,\,+10$) & $0.66\pm0.10$ & $33^{\circ}$ & $2.5\times2.4$ & $126^{\circ}$ \\
B: ($+26,\,+66$) & $0.83\pm0.10$ & $316^{\circ}$ & $2.9\times1.3$ & $41^{\circ}$ \\
C: ($+26,\,+66$) & $2.45\pm0.09$ & $272^{\circ}$ & $2.9\times1.8$ & $127^{\circ}$\\
D: ($+12,\,+26$) & $\sim 0$ & $N/A$ & $3.2\times1.9$ & $\sim-40^{\circ}$\\
\end{tabular}
\tablenotetext{*}{Elliptical gaussian fits of the identified SO$_2$ emission sources.  Offsets are measured in arcseconds from
the center of the continuum emission (RA=$07^{h}22^{m}58.336^{s}$,
Dec=$-25^{\circ}46'03.063''$).
Source size represents the major and minor axis of the ellipse, and its orientation angle is that of the major axis. Angles are given as rotations from north through east. Fits were performed on maps of the 235151.7 MHz line. \label{tab:offsets}}
\normalsize
\end{table}

\vspace{2cm}

\begin{table}

\caption{Summary of LTE and RADEX modeling results \label{table3}} 

\begin{tabular}{lccccc}

\hline
\hline
Source  & $T_{rot}$ & SO$_{2}$ Column Density & $T_{kin}$ & H$_{2}$ Density  & SO$_{2}$ Abundance ( $f_{SO_{2}}$)\\
& (K) & (cm$^{-2}$) & (K) & (cm$^{-3}$) & \\
\hline
A  & $61_{-17}^{+39}$ & $9.2_{-4.4}^{+7.8}\times10^{15}$~cm$^{-2}$ & 50 & $\apprge2$$\times10^{8}$ & $7\times10^{-10}$\\
B  & $110_{-29}^{+63}$ &  $1.9_{-0.6}^{+0.8}\times10^{16}$ & $\sim$ 75 & $\apprge3$$\times10^{7}$ & $2\times10^{-8}$\\
C  &  $69_{-17}^{+34}$ & $1.1_{-0.4}^{+0.7}\times10^{16}$ & $\sim$ 100 & $\apprge3$$\times10^{7}$ & $8\times10^{-9}$\\
D  & $240_{-52}^{+91}$ & $7.6\pm1.0\times10^{15}$\tablenotemark{*} & $\sim$ 220 & $\apprge4$$\times10^{7}$ & $5\times10^{-10}$\\
\hline
\end{tabular}
\tablenotetext{*}{Values for source D suffer from missing extended flux. Using single dish data, average $T_{rot}=97$ K and $N=1.8\times10^{15}$ cm$^{-2}$ assuming that the emitter is an uniform 4" by 4" region.}
\end{table}

\end{document}